\RequirePackage{ifpdf}
\ifpdf 
\documentclass[pdftex]{sigma}
\else
\documentclass{sigma}
\fi

\numberwithin{equation}{section}

\begin{document}

\allowdisplaybreaks

\renewcommand{\PaperNumber}{012}

\FirstPageHeading

\renewcommand{\thefootnote}{$\star$}

\ShortArticleName{Boundary Liouville Theory: Hamiltonian
Description and Quantization}

\ArticleName{Boundary Liouville Theory:\\
Hamiltonian Description and Quantization\footnote{This paper is a
contribution to the Proceedings of the O'Raifeartaigh Symposium on
Non-Perturbative and Symmetry Methods in Field Theory (June
22--24, 2006, Budapest, Hungary). The full collection is available
at
\href{http://www.emis.de/journals/SIGMA/LOR2006.html}{http://www.emis.de/journals/SIGMA/LOR2006.html}}}

\Author{Harald DORN~$^\dag$ and George JORJADZE~$^\ddag$}
\AuthorNameForHeading{H.~Dorn and G.~Jorjadze}

\Address{$^\dag$~Institut f\"ur Physik der
Humboldt-Universit\"at zu Berlin,\\
$\phantom{^\dag}$~Newtonstra{\ss}e 15, D-12489 Berlin, Germany}
\EmailD{\href{mailto:dorn@physik.hu-berlin.de}{dorn@physik.hu-berlin.de}}

\Address{$^\ddag$~Razmadze Mathematical Institute, M.~Aleksidze 1,
0193, Tbilisi, Georgia}
\EmailD{\href{mailto:jorj@rmi.acnet.ge}{jorj@rmi.acnet.ge}}

\ArticleDates{Received October 17, 2006, in f\/inal form December
11, 2006; Published online January 12, 2007}

\Abstract{The paper is devoted to the Hamiltonian treatment of
classical and quantum properties of Liouville f\/ield theory on a
timelike strip in $2d$ Minkowski space. We give a~complete
description of classical solutions regular in the interior of the
strip and obeying constant conformally invariant conditions on
both boundaries. Depending on the values of the two boundary
parameters these solutions may have dif\/ferent monodromy
properties and are related to bound or scattering states. By
Bohr--Sommerfeld quantization we f\/ind the quasiclassical
discrete energy spectrum for the bound states in agreement with
the corresponding limit of spectral data obtained previously by
conformal bootstrap methods in Euclidean space. The full quantum
version of the special vertex operator $e^{-\varphi }$ in terms of
free f\/ield exponentials is constructed in the hyperbolic
sector.}


\Classification{37K05; 37K30; 81T30; 81T40}

\section{Introduction}

In connection with D-brane dynamics in string theory there has
been a renewed interest in Liou\-ville theory with boundaries.
Within the boundary state formalism of conformal f\/ield theory
the complete set of boundary states representing Dirichlet
conditions (ZZ branes) \cite{ZZ} and generalized Neumann
conditions (FZZT branes) \cite{FZZ,T} has been constructed,
including an intriguing relation between ZZ and FZZT branes
\cite{Martinec}.

While FZZT branes naturally arise in the quantization of the
Liouville f\/ield with certain classical boundary conditions, the
set of ZZ branes, counted by two integer numbers, so far has not
found a complete classical counterpart. It seems to us an open
question whether or not all ZZ branes can be understood as the
quantization of a classical set up. Thus motivated we are
searching for a complete treatment of the boundary Liouville
theory relying only on Minkowski space Hamiltonian methods.

A lot of work in this direction has been performed already in the
early eighties by Gervais and Neveu \cite{GN1,GN1+,GN2,GN2+}, see
also \cite{CG}. They restricted themselves to solutions with
elliptic monodromy representing bound states of the Liouville
f\/ield. Obviously such a restriction is not justif\/ied if one
wants to make contact with the more recent results mentioned
above. They also get a~quantization of the parameters
characterizing the FZZT branes \cite{GN2,GN2+}, which is not
conf\/irmed by the more recent investigations \cite{FZZ,T}.

Our paper will be a f\/irst step in a complete  Minkowski space
Hamiltonian treatment of Liouville f\/ield theory on a strip with
independent FZZT type boundary conditions on both sides, which for
a special choice of the boundary and monodromy parameters become
of ZZ type. Concerning the classical f\/ield theory, parts of our
treatment will be close to the analysis of coadjoint orbits of the
Virasoro algebra in~\cite{Balog}. As a new result we consider the
assignment of these orbits to certain boundary conditions. We also
give a unif\/ied treatment of all monodromies, i.e.\ bound states
and scattering states, derive the related symplectic structure,
discuss a free f\/ield parametrization and perform f\/irst steps
to the quantization.

\renewcommand{\thefootnote}{\arabic{footnote}}
\setcounter{footnote}{0}

\section{Classical description}

Let us consider the Liouville equation
\begin{equation}\label{L-eq}
\left(\partial_{\tau}^2-\partial_\sigma^2\right)\varphi
(\tau,\sigma) +{4m^2}\,e^{2\varphi(\tau,\sigma)}=0
\end{equation}
on the strip $\sigma\in(0,\pi),\tau\in\mathbb{R}^1$, where
$\sigma$ and $\tau$ are space and time coordinates, respectively.
Introducing the chiral coordinates $x=\tau+\sigma$, $\bar
x=\tau-\sigma$ and the exponential f\/ield $V=e^{-\varphi}$,
equation~\eqref{L-eq} can be written as
\begin{equation}\label{L-eq1}
V\,\partial^2_{x\bar x}V-
\partial_xV\,\partial_{\bar x}V=m^2,\qquad V>0.
\end{equation}

The conformal transformations of the strip are parameterized by
functions $\xi(x)$, which satisfy the conditions
\begin{equation}\label{conf-tr}
\xi'(x)>0,\qquad \xi(x+2\pi)=\xi(x)+2\pi.
\end{equation}
Note that the function $\xi$ is the same for the chiral and the
anti-chiral coordinates $x \mapsto \xi(x),$ $\bar x \mapsto
\xi(\bar x)$. This group usually is denoted by
$\widetilde{\mbox{Diff}}_+(S^1)$, since it is a covering group of
the group of orientation preserving dif\/feomorphisms of the
circle $\mbox{Diff}_+(S^1)$.

The space of solutions of equation~(\ref{L-eq1}) is invariant
under the transformations
\begin{equation}\label{conf-tr-V}
V(x,\bar x) \mapsto \frac{1}{\sqrt{\xi'(x)\xi'(\bar x)}}
V(\xi(x),\xi(\bar x)),
\end{equation}
which is the basic symmetry of the Liouville model, and a theory
on the strip has to be specif\/ied by  boundary conditions
invariant with respect to (\ref{conf-tr-V}).

The invariance of the Dirichlet conditions
\begin{equation}\label{ZZ}
V|_{\sigma=0}=0=V|_{\sigma=\pi}
\end{equation}
is obvious, but note that the corresponding Liouville f\/ield
becomes singular $\varphi\rightarrow +\infty$ at $\sigma=0$ and
$\sigma=\pi$.

Taking into account that $\xi'(x)=\xi'(\bar x)$ at the boundaries,
another set of invariant boundary conditions can be written in the
Neumann form
\begin{equation}\label{FZZT}
 \partial_\sigma V|_{\sigma=0}=- 2ml,\qquad
\partial_\sigma V|_{\sigma=\pi}=2mr,
\end{equation}
with  constant boundary parameters $l$ and $r$. We study the
Minkowskian case and our aim is to develop the operator approach
similarly to the periodic case \cite{BCT,BCT+, OW, T1, JW}. In
this section we describe the conformally invariant classes of
Liouville f\/ields on the strip and give their Hamiltonian
analysis; preparing, thereby, the systems for quantization.

The energy-momentum tensor of Liouville theory
\begin{equation}\label{T}
T=\frac{\partial^2_{xx} V(x,\bar x)}{V(x,\bar x)} , \qquad \bar T
=\frac{\partial^2_{\bar x\bar x} V(x,\bar x)}{V(x,\bar x)}
\end{equation}
is chiral $\partial_{\bar x}T=0=\partial_{ x}\bar T$.  The linear
combinations $T+\bar T$  and $T-\bar T$ correspond to the energy
density ${\cal E}$ and the energy f\/low $\cal P$, respectively,
\begin{gather*}
{\cal E}=T+\bar T=\frac{1}{2} \left(\partial_\tau\varphi\right)^2
+\frac{1}{2} \left(\partial_\sigma\varphi\right)^2
+{2m^2} e^{2\varphi}-\partial^2_{\sigma\sigma}\varphi ,\nonumber\\
{\cal P}=T-\bar T=\partial_\tau\varphi \partial_\sigma\varphi-
\partial^2_{\tau\sigma}\varphi.
\end{gather*}

The Neumann conditions (\ref{FZZT}) provide vanishing
 energy f\/low at the boundaries, which leads to
\begin{equation}\label{T=T}
T(\tau)=\bar T(\tau)\qquad \mbox{and}\qquad  T(\tau+2\pi)=T(\tau).
\end{equation}
The Dirichlet condition (\ref{ZZ}) allows ambiguities for the
boundary behaviour of $T$ and $\bar T$. In this case we introduce
the conditions (\ref{T=T}) as additional to (\ref{ZZ}), which
means that we assume regularity of $T$ and $\bar T$ at the
boundaries and require vanishing energy f\/low there. Then, due
to~(\ref{T}) and~(\ref{L-eq1}), the f\/ield $V$, in both cases
(\ref{ZZ}) and (\ref{FZZT}), can be represented by
\begin{equation}\label{V}
V(x,\bar x)={m}\left[a\,\psi(\bar x)\psi(x)+b\psi(\bar x)\chi(x)+
c\chi(\bar x)\psi(x)+d\chi(\bar x)\chi(x)\right].
\end{equation}
Here $\psi(x)$, $\chi(x)$ are linearly independent solutions of
Hill's equation
\begin{equation}\label{Hill}
\psi''(x)=T(x) \psi(x) , \qquad \chi''(x)=T(x) \chi(x)
\end{equation}
with the unit Wronskian
\begin{equation}\label{Wronsk}
\psi(x)\chi'(x)-\psi'(x)\chi(x)=1,
\end{equation}
and the coef\/f\/icients $a$, $b$, $c$, $d$ form a
$SL(2,\mathbb{R})$ matrix: $ad-bc=1$. With the notations
\begin{equation*}
\Psi=\left( \begin{array}{cr}
  \psi\\\chi \end{array}\right),\qquad \Psi^T=\left(\psi~~\chi\right),\qquad
A=\left( \begin{array}{cr}
  a&b\\c&d \end{array}\right),
\end{equation*}
equation (\ref{V}) becomes $V(x,\bar x)=m \Psi^T(\bar x) A
\Psi(x)$. The periodicity of $T(x)$ leads to the mo\-nod\-romy
property $\Psi(x+2\pi)=M\Psi(x)$, with $M\in SL(2,\mathbb{R})$.
The Wronskian condition (\ref{Wronsk}) is invariant under the
$SL(2,\mathbb{R})$ transformations $\Psi\mapsto S \Psi$, which
transform the monodromy matrix by $M \mapsto SMS^{-1}$. A special
case is $M=\pm I$, which is invariant under the $SL(2,\mathbb{R})$
maps. Otherwise $M$
 can be transformed  to one of the following forms \cite{Balog}
\begin{gather}
M_h=\pm\left( \begin{array}{cr} e^{-\pi p}&0\\0&e^{\pi
p}\end{array}\right),\quad p>0, \qquad M_p=\pm\left(
\begin{array}{cr}
1&0\\ b&1 \end{array}\right), \quad b=\pm 1,\nonumber\\
  M_e=\pm\left(
\begin{array}{cr}
~\cos \pi\theta&\sin \pi\theta\\-\sin \pi\theta&\cos \pi\theta
\end{array}\right),\quad \theta\in (0,1),\label{M}
\end{gather}
which are called hyperbolic, parabolic and elliptic monodromies,
respectively. The matri\-ces~$A$,~$M$ and the freedom related to
the transformations $\Psi\mapsto S \Psi$ can be specif\/ied by the
boundary conditions. First we consider the case (\ref{ZZ}).

\subsection{Dirichlet condition}

The functions $\psi^2(\tau)$, $\chi^2(\tau)$ and
$\psi(\tau)\chi(\tau)$ are linearly independent and the boundary
condition $V|_{\sigma=0}=0$ with (\ref{V}) leads to $a=0=d=b+c$.
Then, we f\/ind
\begin{equation}\label{V-ZZ}
V(x,\bar x)=m\left[\psi(\bar x)\chi(x)- \chi(\bar
x)\psi(x)\right],
\end{equation}
which corresponds to $b=1=-c$. Note that the  behaviour of
(\ref{V-ZZ}) near to the boundary $\sigma\sim \epsilon$ is given
by $V=2m\epsilon +O(\epsilon^3)$, the other choice $c=1=-b$
corresponds to negative $V$ near to $\sigma=0$ and  has to be
neglected. Applying the boundary condition $V|_{\sigma=\pi}=0$ to
(\ref{V-ZZ}) and using the monodromy property we obtain
$\Psi(x+2\pi)=\pm \Psi(x)$. Expanding $V$ now near to the right
boundary $\sigma\sim\pi-\epsilon$, we get $V=\mp2m\epsilon
+O(\epsilon^3)$. Thus, the allowed monodromy is
$\,\Psi(x+2\pi)=-\Psi(x)$ only. After f\/ixing the matrices $A$
and $M$ we have to specify the class of functions $\psi$ and
$\chi$, which ensures the positivity of $V$ in the whole bulk
$\sigma\in (0,\pi)$.

Representing $(\psi ,\chi ) $ in polar coordinates, due to the
unit Wronskian condition, the radial coordinate is f\/ixed in
terms of the angle variable $\xi (x)/2$ resulting in
\begin{equation}\label{psi-chi-ZZ}
\psi(x)=\sqrt{\frac{2}{\xi'(x)}} \cos\frac{\xi(x)}{2},\qquad
\chi(x)=\sqrt{\frac{2}{\xi'(x)}} \sin\frac{\xi(x)}{2},
\end{equation}
and by (\ref{V-ZZ})  the  $V$-f\/ield becomes
\begin{equation}\label{V-xi-ZZ}
V=\frac{2m}{\sqrt{\xi'(x)\,\xi'(\bar x)}}
\sin{\frac{1}{2}\left[\xi(x)-\xi(\bar x)\right]}.
\end{equation}
The obtained monodromy of $\Psi(x)$ leads to
$\xi(x+2\pi)=\xi(x)+2\pi(2n+1)$ with arbitrary integer~$n$, but
(\ref{V-xi-ZZ}) is positive in the whole  strip
 $\sigma\in(0,\pi)$
for $n=0$ only. Thus, $\xi(x)$ turns out to be just a function
parameterizing a dif\/feomorphism  according to (\ref{conf-tr}).
Equation (\ref{V-ZZ}) is invariant under the $SL(2,\mathbb{R})$
transformations $\Psi\mapsto S\Psi$. The corresponding
inf\/initesimal form of $\xi(x)$ is
\begin{equation}\label{SL(2)}
\xi(x)~\mapsto~\xi(x)+\varepsilon_1+\varepsilon_2\cos\xi(x)
+\varepsilon_3\sin\xi(x)
\end{equation}
and the space of solutions (\ref{V-xi-ZZ}) is identif\/ied with
$\widetilde{\mbox{Diff}}_+(S^1)/\widetilde{SL}(2,\mathbb{R})$.

The energy momentum tensor (\ref{T}) calculated from
(\ref{V-xi-ZZ}) reads
\begin{equation}\label{T-ZZ}
T(x)=-\frac{1}{4} \xi'\,^2(x)+S_\xi(x),\qquad \mbox{with}\qquad
S_\xi(x)=\left(\frac{\xi''(x)}{2\xi'(x)}\right)^2-
\left(\frac{\xi''(x)}{2\xi'(x)}\right)',
\end{equation}
and for $\xi(x)=x$ it is constant $T=-\frac{1}{4}$. The
corresponding Liouville f\/ield is time-independent
\begin{equation}\label{ZZ-vacuum}
\varphi_0=-\log( 2m\sin\sigma ),
\end{equation}
and it is associated with the vacuum of the system. The vacuum
solution is invariant under the $SL(2,\mathbb{R})$ subgroup of
conformal transformations generated by the vector f\/ields
$\partial_x$, $\cos x \partial_x$ and $\sin x \partial_x$ and this
symmetry is a particular case of (\ref{SL(2)}) for $\xi (x)=x$.
Therefore, the solutions of the Liouville equation with Dirichlet
boundary conditions form the conformal orbit of the vacuum
solution (\ref{ZZ-vacuum}). The energy functional
\begin{equation}\label{E}
E= \int_0^{\pi}d\sigma  \, (T(\tau +\sigma )+T(\tau -\sigma ))=
\int_0^{2\pi}dx  \, T(x)
\end{equation}
on this orbit is bounded below and the minimal value is achieved
for the vacuum conf\/iguration~\cite{Balog}.

To get the Hamiltonian description we f\/irst specify the boundary
behaviour of Liouville f\/ields. Due to (\ref{Hill}),
(\ref{Wronsk}), (\ref{V-ZZ}), near to the boundaries $V$ is given
by
\begin{equation*}
V=2m\epsilon+\frac{4m}{3} T(\tau) \epsilon^3+O(\epsilon^5)
\end{equation*}
and we f\/ind
\begin{alignat}{3}\label{phi-pi-b}
&\varphi=-\log2m\epsilon-\frac{2}{3} T  \epsilon^2+ O(\epsilon^4)
,\qquad && \partial_\tau\varphi =
-\frac{2}{3} \dot T\epsilon^2+O(\epsilon^4),& \\
\label{phi'-b}
&\partial_\sigma\varphi=\pm\left(\frac{1}{\epsilon}+
\frac{4}{3}T\epsilon\right)+ O(\epsilon^3),\qquad&&
\partial^2_{\sigma\sigma}\varphi=\frac{1}{\epsilon^2}-
\frac{4}{3}T+O(\epsilon^2).&
\end{alignat}
The signs $+$ and $-$ for $\partial_\sigma\varphi$ correspond to
the right ($\sigma=\pi$) and left ($\sigma=0$) boundaries and the
argument of $T$ is $\tau +\pi $ and $\tau$, respectively. The
Liouville equation (\ref{L-eq}) is equivalent to the Hamilton
equations obtained from the canonical action
\begin{equation}\label{H-ZZ}
S=\int d\tau\int_0^\pi d\sigma\left[ \pi\,\dot\varphi-
\left(\frac{1}{2}
\pi^2+\frac{1}{2}\left(\partial_\sigma\varphi\right)^2 +2m^2
e^{2\varphi}-\partial^2_{\sigma\sigma}\varphi\right) \right],
\end{equation}
with the Hamiltonian given by the energy functional (\ref{E}).
Note that $\partial^2_{\sigma\sigma}\varphi$ can not be integrated
into a boundary term due to the singularities (\ref{phi'-b}).

The canonical 2-form related to (\ref{H-ZZ})
\begin{equation}\label{omega-can}
\omega=\int_0^\pi d\sigma \delta\pi(\tau,\sigma)\wedge
\delta\varphi(\tau,\sigma)
\end{equation}
is well def\/ined on the class of singular functions
(\ref{phi-pi-b}) and using the parameterization
(\ref{V-ZZ})--(\ref{psi-chi-ZZ}), we f\/ind this 2-form in terms
of the $\xi$-f\/ield (see Appendix A)
\begin{equation}\label{omega-ZZ}
\omega= \frac{1}{4}\int_0^{ 2\pi}dx \left[
\frac{\delta\xi''(x)\wedge\delta\xi'(x)}{\xi'^2(x)}
-\delta\xi'(x)\wedge\delta\xi(x)\right].
\end{equation}
It  is degenerated, but has to be reduced on the space
$\widetilde{\mbox{Diff}}_+(S^1)/\widetilde{SL}(2,\mathbb{R})$,
where it becomes symplectic.

One can of course also study the case with Dirichlet conditions on
one and Neumann conditions on the other boundary of the strip, say
\begin{equation*}
V\vert _{\sigma   =0}=0,\qquad
\partial_{\sigma}V\vert_{\sigma=\pi}=2mr.
\end{equation*}
Starting again with (\ref{V-ZZ}), the boundary condition at
$\sigma =\pi$, together with the unit Wronskian forces the trace
of the monodromy matrix to be equal to $2r$. Thus for $\vert r
\vert <1$ one has elliptic and for $r >1$ hyperbolic monodromy
($r<-1$ is excluded by arguments of the next section).  Then an
analysis similar to the one above gives for $-1\leq r < 1$
\begin{equation*}
V=\frac{2m}{\theta\sqrt{\xi'(x) \xi'(\bar x)}}
\sin{\frac{\theta}{2}\left[\xi(x)-\xi(\bar x)\right]},
\end{equation*}
with $r=\cos\pi\theta$. The related energy momentum tensor is
\begin{equation*}
T(x)=-\frac{\theta ^2}{4}\xi'^2(x)+S_\xi(x).
\end{equation*}
The result for $r>1$ is obtained by the replacement $\theta =ip$
with real~$p$.

\subsection{Generalized Neumann conditions}

To analyze the Neumann conditions (\ref{FZZT}) we f\/irst
construct the f\/ields corresponding to constant energy-momentum
tensor $T(x)=T_0$ and then obtain others by conformal
transformations. It appears that this construction covers all
regular Liouville f\/ields on the strip.

For $~T_0=p^2/4 >0~$ convenient solutions of Hill's equation are
\begin{equation*}
\psi(x)=\cosh(px/2),\qquad \chi(x)=\frac{2}{p}\sinh(px/2).
\end{equation*}
The related monodromy is hyperbolic. To get it in the normal form
(\ref{M}) one has to switch to corresponding exponentials. The
form chosen allows a smooth limit for $\psi$ and $\chi $ to the
parabolic and elliptic cases below. The corresponding $V$-f\/ield
(\ref{V}), which obeys the Neumann conditions~(\ref{FZZT}), reads
\begin{equation}\label{V-h}
V_0(x,\bar x)=\frac{2\,m}{p\sinh \pi p} \big (u\cosh
{p(\tau-\tau_0)}+ l\cosh{p(\sigma-\pi)} +r \cosh {p\sigma}\big ),
\end{equation}
where
\begin{equation}\label{l-r,u}
 u=u(l,r;p) =\sqrt{l^2+r^2
+2l r \cosh\pi p+\sinh^2\pi p},
\end{equation}
and $\tau_0$ is an arbitrary constant. The positivity of the
$V$-f\/ield (\ref{V-h}) imposes restrictions on the parameters of
the theory. One can show that if $l< -1$ or $r< -1$, then the
positivity of (\ref{V-h}) fails for any $p>0$; but if $l\geq -1$
and $r\geq -1$, then (\ref{V-h}) is positive in the whole bulk
$\sigma\in (0,\pi)$ for all $p>0$.

For $T_0=0$  we can choose $\psi(x)=1$, $\chi(x)=x$ and obtain
\begin{equation}\label{parab}
V_0=\frac{m}{\pi} \left[(l+r)
\left(\sigma^2-(\tau-\tau_0)^2\right) -\pi l(2\sigma-\pi) -\pi^2
\frac{1+lr}{l+r}\right].
\end{equation}
This case corresponds to the parabolic monodromy. The positivity
of (\ref{parab}) requires $l\geq -1$, $r\geq -1$ and $l+r<0$.
Note that (\ref{parab}) is also obtained from (\ref{V-h}) in the
limit $p\rightarrow 0$, if $l+r<0$. Among these parabolic
solutions there are two time-independent solutions
\begin{equation*}
V_0=2m\sigma\qquad \mbox{and}\qquad V_0=2m(\pi-\sigma),
\end{equation*}
which correspond to the degenerated cases of (\ref{parab}) for
$l=-1$, $r=1$ and $l=1$,  $r=-1$, respectively.

For $T_0=-\theta ^2/4 <0$  the pair
\begin{equation*}
\psi(x)=\cos(\theta x/ 2),\qquad \chi(x)=\frac{2}{\theta}
\sin(\theta x/2),
\end{equation*}
has elliptic monodromy and the $V$-f\/ield becomes
\begin{equation}\label{V-e}
V_0=-\frac{2\, m}{\theta\sin \pi\theta} \big (u \cos
{\theta(\tau-\tau_0)}+ l\cos{\theta(\sigma-\pi)}+r\cos
{\theta\sigma} \big ) ,
\end{equation}
with $u = u(l,r;i\theta ) =\sqrt{l^2+r^2 +2l r \cos \pi
\theta-\sin^2 \pi\theta}$. Due to oscillations in $\sigma$ the
positivity of~(\ref{V-e}) in the whole bulk fails if $\theta>{1}$;
hence, $\theta\leq {1}$.

Other restrictions on the parameters come from the analysis of the
equation
\begin{equation}\label{ellipse}
l^2+r^2 +2l r \cos\pi\theta-\sin^2\pi\theta=0,
\end{equation}
which def\/ines an ellipse on the  $(l,r)$-plane. The ellipse is
centered at the origin, its half axis with length  $\sqrt{1\pm
  \cos\pi\theta}$ are situated on the
lines $r\pm l=0$. It is tangential to the lines $l=-1$ and $r=-1$
at the points $B$ and $C$ with the coordinates $B=(-1,
\cos\pi\theta)$ and $C=(\cos \pi\theta, -1)$ . The curve $BC$ in
Fig.~1 indicates the corresponding arc of the ellipse. The
positivity of $V$-f\/ield (\ref{V-e}) requires $(l,r)\in
\Omega_{ABC}$, where $\Omega_{ABC}$ is the `triangle' bounded by
the lines $l=-1$, $r=-1$ and the arc   $BC$.

\begin{figure}[t]
\centerline{\includegraphics[width=7.5cm]{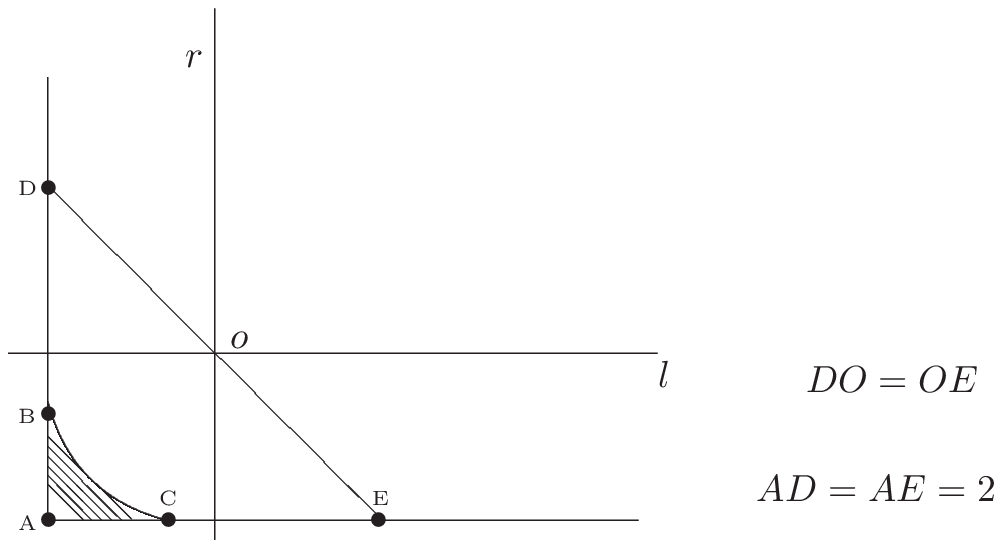}}

\vspace{-2mm} \caption{}
\end{figure}

Having so far discussed which values of $l$ and $r$ are allowed
for given $\theta$, we now turn the question around. We start with
$(l,r)$ somewhere in the triangle $ADE$, then all $\theta $
obeying
\[
0\leq \theta\leq \theta_*(l,r)
\]
with $\theta_*(l,r)$ def\/ined by
\begin{equation}\label{theta-0}
\cos\pi\theta_*=-lr+\sqrt{(1-l^2)(1-r^2)},\qquad
0<\theta_*\leq{1}.
\end{equation}
are allowed\footnote{Note that (\ref{theta-0}) is the larger root
of equation (\ref{ellipse}) for $\cos\pi\theta$. The smaller root
corresponds to the case when the point $(l,r)$ is  on the ellipse
but does not belong to the arc $BC$.}. For $\theta =\theta_*$ the
$V$-f\/ield is time-independent
\begin{equation}\label{V-0}
V_0=-\frac{2\,m}{\theta_*\sin\pi\theta_*} \left[
l\cos{\theta_*(\sigma-\pi)} +r\cos{\theta_*\sigma}\right].
\end{equation}
For $\theta_*=1$, the arc $BC$ degenerates to the point $A$ and
the Liouville f\/ield (\ref{V-0}) becomes the Dirichlet  (ZZ)
vacuum (\ref{ZZ-vacuum}). Thus, for the critical values of the
boundary parameters $l=r=-1$ the Neumann (FZZT) case contains the
$ZZ$ case.

The above analysis shows that the admissible values of the
boundary parameters are $ l\geq-1$, $ r\geq-1$ and for a given
$(l, r)$ obeying these constraints there are the following
restrictions on $T_0$
\begin{alignat*}{6}
& l+r>0 && && && \Rightarrow && \quad T_0>0,& \nonumber\\
& l+r=0\quad && \mbox{and}\quad &&  l\neq \pm 1 &&\Rightarrow && \quad T_0>0,&\nonumber\\
&l+r=0 && \mbox{and}&& l=\pm 1 \quad && \Rightarrow && \quad T_0\geq 0,& \nonumber\\
& l+r<0 && && && \Rightarrow && \quad T_0\geq -\frac{\theta _*^2(l,r)}{4}.&
\end{alignat*}
Thus, we have all three monodromies if $(l,r)$ is inside the
triangle $ADE$, and only hyperbolic monodromy if $(l,r)$ is
outside of it. In the f\/irst case (\ref{V-e}) is a continuation
of (\ref{V-h}) from positive to negative $T_0$ and for $T_0=0$ it
coincides with (\ref{parab}).

{\samepage The conformal orbits, generated out of these f\/ields with
constant $T(x)=T_0$ can be written as
\begin{gather}
V(x,\bar x)=\frac{m\left(\xi'(x)\xi'(\bar
x)\right)^{-\frac{1}{2}}}{p\sinh \pi p} \Big (u(l,r;p)
e^{-(\xi(x)+\xi(\bar x))p/2}
+ u(l,r;p) \,e^{(\xi(x)+\xi(\bar x))p/2}\nonumber\\
\phantom{V(x,\bar x)=}{}+(l\,e^{-\pi p }+r )
 e^{(\xi(x)-\xi(\bar x))p/2}+
 (l\,e^{\pi p}+r)
 e^{(\xi(\bar x)-\xi(x))p/2}
\Big ),\label{V-xi}
\end{gather}
where $p=2\sqrt{T_0}$ and $\xi(x)$ is the group parameter on the
orbits. The parameter $\tau_0$ is absorbed by the zero mode of
$\xi(x)$.}

The energy-momentum tensor for (\ref{V-xi}) is given by
\begin{equation}\label{FZZT-T}
T(x)=T_0 \xi'^2(x)+S_\xi(x).
\end{equation}
Here $S_\xi(x)$ is the Schwarz derivative (\ref{T-ZZ}), which
def\/ines the inhomogeneous part of the transformed $T(x)$. Note
that there are other orbits, either with $T_0<-\frac{1}{4}$,
 or those, which do not contain orbits with constant $T(x)$ \cite{Balog}.
Using the classif\/ication of Liouville f\/ields by coadjoint
orbits, one can show that the orbits (\ref{V-xi}) with
$T_0\geq-\frac{1}{4}$ cover all regular Liouville f\/ields on the
strip. One can also prove that the energy functional (\ref{E}) is
bounded below just on these orbits only~\cite{Balog}. Thus,
boundary Liouville theory selects the class of f\/ields with
bounded energy functional and, therefore, its quantum theory
should provide  highest weight representations of the Virasoro
algebra.

The Hamiltonian approach based on the action (\ref{H-ZZ}) is
applicable for the Neumann conditions as well. Indeed, the
boundary conditions (\ref{FZZT}) are equivalent to
\begin{equation*}
\left(\partial_\sigma\varphi- 2ml e^\varphi\right)|_{\sigma=0}=0=
\left(\partial_\sigma\varphi+ 2mr e^\varphi\right)|_{\sigma=\pi}
\end{equation*}
and its variation yields $\partial_\sigma(\delta\varphi)-
(\partial_\sigma\varphi)\delta\varphi|_{\sigma=0,\pi}=0$, which
cancels the boundary term for the variation of (\ref{H-ZZ}). The
Legendre transformation of (\ref{H-ZZ}) and the integration of
$\partial_{\sigma\sigma}^2\varphi$ into the boundary terms leads
to the action \cite{FZZ}
\begin{equation}\label{S-FZZ}
S=\int d\tau\int_0^\pi d\sigma\frac{1}{2}
\left[(\partial_\tau\varphi)^2-(\partial_\sigma\varphi)^2-
4m^2e^{2\varphi}\right]-2m\int d\tau \left[le^{\varphi(\tau,0)}+ r
e^{\varphi(\tau,\pi)}\right].
\end{equation}
Note that for $l=-1$ or/and $r=-1$ the $V$-f\/ield vanishes at the
boundary for $\tau=\tau_0$ and one can not pass to (\ref{S-FZZ}),
due to the singularities of $e^{\varphi}$.

The canonical 2-form (\ref{omega-can}) can be calculated in the
variables $(T_0 ,\xi)$ similarly to (\ref{omega-ZZ}) and we obtain
(see Appendix~A)
\begin{gather}\label{omega-FZZT}
\omega= \delta T_0 \wedge\int_0^{2\pi}dx\xi'(x)\delta\xi(x)+ T_0
\int_0^{2\pi}dx\delta\xi'(x)\wedge\delta\xi(x)+
\frac{1}{4}\int_0^{ 2\pi}dx
\frac{\delta\xi''(x)\wedge\delta\xi'(x)}{\xi'^2(x)}.
\end{gather}
In the context of Liouville theory this symplectic form was
discussed in \cite{ASh}, where it was obtained as a generalization
of the symplectic form on the co-adjoint orbits of the $2d$
conformal group. Note that the form of  (\ref{omega-FZZT}) does
not depend on the boundary parameters $l$ and $r$. This dependence
implicitly is encoded in the domain of $T_0$: if $(l,r)$ is inside
the triangle $ADE$, then $T_0 \geq -\theta _*^2$,
(\ref{theta-0}); and $T_0>0$ if $(l,r)$ is outside the triangle.
The symplectic form (\ref{omega-FZZT}) provides the following
Poisson brackets
\begin{equation}\label{PB,s-xi-s}
\{T_0,\,\xi(x)\}=\frac{1}{2\pi} , \qquad
\{\xi(x),\,\xi(y)\}=\frac{1}{4T_0}\left(\frac{\sinh\left(
2\sqrt{T_0}\,\lambda(x,y)\right)}{\sinh\left(2\pi\sqrt{T_0}\,\right)}
-\frac{\lambda(x,y)}{\pi}\right),
\end{equation}
where $\lambda(x,y)=\xi(x)-\xi(y)-\pi\epsilon(x-y)$ and
$\epsilon(x)$ is the stair-step function: $\epsilon(x)=2n+1$, for
$x\in(2\pi n, 2\pi n+2\pi)$, which is related to the periodic
$\delta$-function by $\epsilon'(x)=2\delta(x)$.

From (\ref{PB,s-xi-s}) we obtain
\begin{gather*}
\{T(x),\xi(y)\}=\xi'(x)\delta(x-y),\\
\{T(x),T(y)\}=T'(y)\delta(x-y)-2T(y) \delta'
(x-y)+\frac{1}{2}\delta'''(x-y),
\end{gather*}
which def\/ine the conformal transformations for the f\/ields
$\xi(x)$ and~$T(x)$.

Using the Fourier mode expansion for $\xi(x)$
\begin{equation*}
\xi(x)=x+\sum_{n\in Z}{\xi_n} e^{-inx}
\end{equation*}
and the f\/irst equation of (\ref{PB,s-xi-s}), we f\/ind that
$2\pi\xi_0$ is the canonical conjugated to $T_0$
\begin{equation}\label{PB,s-xi}
 \{T_0,2\pi\xi_n\}=\delta_{n\,0}.
\end{equation}
For $T_0<0$ the variable  $\alpha=2\sqrt{-T_0} \xi_0$ is cyclic
$(\alpha \sim \alpha+2\pi)$, since the exponentials in
(\ref{V-xi}) become oscillating. By (\ref{PB,s-xi}), $\alpha$ is
canonical conjugated to
 $\pi\theta $, where $\theta=2\sqrt{-T_0}$.

This allows a remarkable f\/irst conclusion concerning
quantization. Semi-classical Bohr--Sommerfeld quantization of
$\theta $ yields $\theta _n = -\hbar n/\pi + \theta _*(l,r)$,
which implies a quantization of~$T_0$
\begin{gather}\label{semi}
(T_0)_n =-\frac{1}{4}\theta _n ^2 = -\frac{\hbar ^2}{4\pi^2}
n^2+\frac{\hbar}{2\pi}\theta _*(l,r)n-\frac{1}{4}\theta _*^2 ,
\end{gather}
with integer $n$ as long as $(T_0)_n<0$. As shown in Appendix B,
this spectrum with the identif\/ication $\hbar =2\pi b^2$ and the
trivial shifts $n^2\mapsto n(n+1)$, $(T_0)_n\mapsto (T_0)_n+1/4$
agrees with the quasiclassical expansion of the spectrum derived
in \cite{T} by highly dif\/ferent methods.

After this short aside we start preparing for the full
quantization of our system. The variables $(\xi,  p   )$ are not
suitable for  this purpose due to  the    complicated form of the
Poisson brackets~(\ref{PB,s-xi-s}). A natural approach in this
direction is a free-f\/ield parameterization with a perspective of
cano\-ni\-cal quantization.

For $T_0=p^2/4 >0$, free-f\/ield variables can be introduced
similarly to the periodic case \cite{JW}
\begin{equation}\label{ff}
\phi(x)=\frac{p\xi(x)}{2}+\frac{1}{2}\log\xi'(x)-
\frac{1}{2}\log\frac{m\,u(l,r;p)}{p\sinh \pi p}.
\end{equation}
Here the $x$-independent part given by the last term is chosen for
further convenience, for   $u(l,r;p)$ see (\ref{l-r,u}). The
f\/ield $\phi(x)$ obviously has the monodromy
$\phi(x+2\pi)=\phi(x)+\pi p$, which allows the mode expansion
\begin{equation*}
\phi(x)=\frac{q}{2\pi}+\frac{px}{2}+
\frac{i}{\sqrt{4\pi}}\sum_{n\neq 0}\frac{a_n}{n} e^{-inx}.
\end{equation*}
The integration of (\ref{ff}) yields
\begin{equation}\label{xi-phi}
\xi(x)=\frac{1}{p}\log \frac{m u A_p(x)}{2\sinh^2\pi p},
\end{equation}
where $A_p(x)$ is the integral of the equation $A_p'(x)= 2\sinh\pi
p e^{2\phi(x)}$ with the monodromy property $A_p(x+2\pi)=e^{2\pi
p}  A_p(x)$ and it can be written as
\begin{equation*}
A_p(x)=\int_0^{2\pi}dy e^{2\phi(x+y)-\pi p}.
\end{equation*}

The free-f\/ield form of (\ref{omega-FZZT})
\begin{equation*}
\omega=\int_0^{2\pi}dx \delta\phi'(x)\wedge\delta\phi(x)+ \delta
p\wedge\delta\phi(0)=\delta p\wedge\delta
q+\frac{1}{2i}\sum_{n\neq 0} \frac{1}{n} \delta a_n\wedge \delta
a_{-n},
\end{equation*}
follows from the direct computation and it provides the canonical
brackets{\samepage
\begin{equation}\label{PB,a_n}
\{\phi(x),\phi(y)\}=\frac{1}{4} \epsilon(x-y),\qquad
\mbox{or}\qquad \{p,q\}=1,\qquad \{a_n,a_m\}=in\delta_{n+m,0}.
\end{equation}
Note that these brackets and (\ref{xi-phi}) lead
to~(\ref{PB,s-xi-s}).}

The energy-momentum (\ref{FZZT-T}) takes also a free-f\/ield form
with a linear  improvement  term
\begin{equation}\label{T-phi}
T(x)=\phi\,'^2(x)-\phi''(x),
\end{equation}
and by (\ref{PB,a_n}) we have
\begin{equation}\label{PB,T-phi'}
\{T(x), \phi'(y)\}=\phi''(y)\delta(x-y)-\phi'(y)\delta'(x-y)
+{1}/{2}\delta''(x-y).
\end{equation}

Inserting (\ref{xi-phi}) into (\ref{V-xi}), we f\/ind
\begin{equation}\label{V-phi}
V=e^{-[\phi(x)+\phi(\bar x)]}\left[1+mb_pA_p(x)+ mc_pA_p(\bar x)+
m^2d_pA_p(x)A_p(\bar x)\right],
\end{equation}
with
\begin{equation}\label{b-p,c-p}
b_p=\frac{le^{-\pi p}+r}{2\sinh^2 \pi p},\qquad c_p=\frac{le^{\pi
p}+r}{2\sinh^2\pi p},\qquad d_p=\frac{u^2(l,r;p)}{4\sinh^4\pi p}.
\end{equation}
The f\/ield $\Phi=\phi(x)+\phi(\bar x)$ is the full free-f\/ield
on the strip. It satisf\/ies  for all allowed values of $l$ and
$r$ the standard Neumann boundary conditions
$\partial_\sigma\Phi|_{\sigma=0}
=0=\partial_\sigma\Phi|_{\sigma=\pi}$ and has the following mode
expansion
\begin{equation*}
\Phi(\tau,\sigma)=\frac{q}{\pi}+{p\tau}+
\frac{i}{\sqrt{\pi}}\sum_{n\neq 0}\frac{a_n}{n} e^{-in\tau}\,\cos
n\sigma,
\end{equation*}
Since $p>0$, $A_p(x)$ and $A_p(\bar x)$ vanish for
$\tau\rightarrow -\infty$. Therefore $\Phi(\tau,\sigma)$ is the
$in$-f\/ield for the Liouville f\/ield:
$\varphi(\tau,\sigma)\rightarrow \Phi(\tau,\sigma)$, for
$\tau\rightarrow -\infty$.

The chiral $out$-f\/ield is introduced similarly to (\ref{ff})
replacing $p$ by $-p$ and its mode expansion can be written as
\begin{equation*}
\phi_{\rm out}(x)=\frac{\tilde q}{2\pi}-\frac{px}{2}+
\frac{i}{\sqrt{4\pi}}\sum_{n\neq 0}\frac{\tilde a_n}{n} e^{-inx},
\end{equation*}
The relation between $in$ and $out$ f\/ields
\begin{equation*}
\phi_{\rm out}(x)=\phi(x)-\log \frac{muA_p(x)}{2\sinh^2\pi p},
\end{equation*}
def\/ines a canonical map between the modes $(p,q; a_n)$
 and $(\tilde q, -p, \tilde a_n)$. Quantum mechanically this map
is given by the $S$-matrix and f\/inding its closed form is one of
the basic open problems of Liouville theory.

\section{Canonical quantization}

In this section we consider canonical quantization applying the
technique developed for the periodic case \cite{BCT,BCT+, OW, T1,
JW}. Our discussion has some overlap with \cite{CG}. But in
contrast to their parametrization in terms of two related free
f\/ields we use only one parametrizing free f\/ield. We mainly
treat the hyperbolic case. The quantum theory of other sectors can
be obtained by analytical continuation in the zero mode $p$,
choosing appropriate values of the boundary parameters $(l,r)$.

The canonical commutation relations
\begin{equation*}
[q,p]=i\hbar,\qquad [a_m,a_n^*]=\hbar m \delta_{mn}\qquad (m>0,\
n>0),
\end{equation*}
are equivalent to the chiral  commutator
\begin{equation}\label{phi-phi^}
[\phi(x), \phi(y)]=-\frac{i\hbar}{4} \epsilon(x-y),
\end{equation}
and have a standard realization in the Hilbert space
$L^2(\mathbb{R}_+)\otimes\cal{F}$, where $L^2(\mathbb{R}_+)$
corresponds to the momentum representation of the zero modes with
$p>0$ and $\cal{F}$ stands for the Fock space of the non-zero
modes $a_n$. We use the same notations for classical and
corresponding normal ordered quantum expressions, which, in
general, have to be deformed in order to preserve the symmetries
of the theory. The guiding principle for the construction of
quantum operators are the conformal symmetry and inf\/inite
dimensional translation symmetry generated by $\phi'(x)$.
A~semi-direct product of these symmetry groups is provided by the
Poisson bracket (\ref{PB,T-phi'}), which quantum mechanically
admits a deformation of the central term. This implies a
deformation of the coef\/f\/icient in front of the linear term in
the energy-momentum tensor (\ref{T-phi})
\begin{equation*}
T(x)=\phi'^2(x)-\eta \phi''(x).
\end{equation*}
The related Virasoro generators satisfy the standard commutation
relations with the central charge $c=1+{12\pi\eta^2}/{\hbar}$. The
deformation parameter $\eta$ is f\/ixed by conformal properties of
free-f\/ield exponentials. Using the decomposition
$\phi(x)=\phi_0(x)+\phi_+(x)+\phi_-(x)$, with
\begin{equation*}
\phi_0(x)=\frac{q}{2\pi}+\frac{px}{2},\qquad
\phi_+(x)=-\frac{i}{\sqrt{4\pi}} \sum_{n> 0}\frac{a_n^*}{n}
e^{inx},\qquad \phi_-(x)=\frac{i}{\sqrt{4\pi}} \sum_{n>
0}\frac{a_n}{n} e^{-inx},
\end{equation*}
a free-f\/ield exponential is introduced in a standard normal
ordered form
 \begin{equation*}
e^{2\lambda\phi(x)}= e^{2\lambda\phi_0(x)} e^{2\lambda\phi_+(x)}
e^{2\lambda\phi_-(x)}.
\end{equation*}
Requiring unit conformal weight of $e^{2\phi(x)}$, one f\/inds
$\eta=1+b^2$, with $2\pi b^2=\hbar$.

Our aim is to construct the vertex operator corresponding to the
Liouville exponential (\ref{V-phi}). Building blocks for this
construction are the chiral operators
\begin{gather}\label{psi^}
\psi(x)= e^{-\phi(x)}, \\ \label{A^}
A_p(x)=\int_0^{2\pi}dz~e^{2\phi_0(x+z)-\pi p} e^{2\phi_+(x+z)}
e^{2\phi_-(x+z)},\\ \label{chi^} \chi(x) = \psi(x)  A_p(x).
\end{gather}
The operators $\psi(x)$ and $A_p(x)$  are obviously hermitian and
the $p$-dependent shift of $\phi_0$ in (\ref{A^}) is motivated by
hermiticity of  $\chi(x)$ (see (\ref{psi-A^})). The unit conformal
weight of $e^{2\phi(x)}$ provides zero conformal weight of
$A_p(x)$ and, therefore the conformal weights of the operators
$\psi$ and $\chi$ are the same, like in the classical case.
Exchange relations of these operators and their classical
counterparts are derived in Appendix~C. It is important to note
that these relations for the $\psi$ and $\chi$ f\/ields are the
same
\begin{gather}\label{psi-psi^}
\psi(x) \psi(y)=e^{-{i(\hbar}/{4}) \epsilon(x-y)} \psi(y)\psi(x),
\\ \label{chi-chi^}
\chi(x)\chi(y)=e^{-{i(\hbar}/{4})\epsilon(x-y)}\chi(y)\chi(x).
\end{gather}

Based on (\ref{V-phi}), we are looking for the vertex operator $V$
in the form
\begin{gather*}
V(x,\bar x)=e^{{-i(\hbar}/{8})}\left[\psi(\bar x)\psi(x)+B_p
\psi(\bar x)\chi(x)+C_p\chi(\bar x)\psi(x)+D_p \chi(\bar
x)\chi(x)\right],
\end{gather*}
with $p$-dependent coef\/f\/icients $B_p$, $C_p$ and $D_p$. The
phase factor $e^{{-i(\hbar}/{8})}$ provides hermiticity of the
f\/irst term of $V$-operator, which corresponds to the
$in$-f\/ield exponential. The last term describes the
$out$-f\/ield exponential, respectively.

To f\/ix $B_p$, $C_p$ and $D_p$  we use the conditions of locality
and hermiticity
\begin{gather}\label{local-herm}
[V(\tau+\sigma,\tau-\sigma),
V(\tau+\sigma',\tau-\sigma')]=0,\qquad V^*(x,\bar x)=V( x, \bar
x).
\end{gather}
The analysis of these equations can be done ef\/fectively with the
help of  exchange relations between the $\psi$ and $\chi$
operators. There are two kind of exchange relations. The f\/irst
exchanges the ordering of the arguments $x$ and $y$
\begin{gather}
\chi(x)\psi(y)= e^{{i(\hbar}/{4}) \epsilon(x-y)} \!\left[
\frac{\sinh\left(\pi p+{i\hbar}/{2}\right)} {\sinh\pi p}
\psi(y)\chi(x) -i\sin({\hbar}/{2}) \frac{e^{\pi p\epsilon(x-y)}}
{\sinh\pi p} \chi(y)\psi(x)\right]\!,\!\label{chi-psi^}
\end{gather}
and another the ordering of $\chi$ and $\psi$ f\/ields
\begin{gather}
\chi(x) \psi(y)=e^{{i(\hbar}/{4}) \epsilon(x-y)}\nonumber\\
\phantom{\chi(x) \psi(y)=}{}\times\left[ \frac{\sinh\pi p}
{\sinh\left(\pi p-{i\hbar}/{2}\right)}\psi(y)\chi(x)
-i\sin({\hbar}/{2}) \frac{e^{(\pi p-i\hbar/4)\epsilon(x-y)}}
{\sinh\left(\pi p-{i\hbar}/{2}\right)}
\psi(x)\chi(y)\right]\!.\!\label{chi-psi^1}
\end{gather}
Applying these relations to (\ref{local-herm}) we obtain a set of
equations for the functions $B_p$, $C_p$, $D_p$. They relate the
values of these coef\/f\/icients with shifted arguments and we
have found the following solution of these equations (see Appendix
D)
\begin{gather}\label{b_p}
B_p=m_b \frac{l_be^{-(\pi p-i\hbar/2)}+r_b}{2\sinh\pi p \sinh(\pi
p-i\hbar/2)},
\\
\label{c_p} C_p=m_b \frac{l_be^{(\pi p+i\hbar/2)}+r_b}{2\sinh\pi p
\sinh(\pi p+i\hbar/2)},
\\
\label{d_p} D_p=\frac{m^2_b}{4\sinh\pi p \sinh(\pi p+i\hbar)}
\left(1+ \frac{l_b^2+r_b^2+2l_br_b\cosh(\pi p+i\hbar/2)}
{\sinh^2(\pi p+i\hbar/2)}\right).
\end{gather}
The parameters $m_b$, $l_b$ and $r_b$ arise in the solution as $p$
independent constants. Comparing these expressions with their
classical analogs (\ref{b-p,c-p}), we f\/ind a naturally
interpretation of $m_b$ and ($l_b,r_b$) as a renormalized mass and
renormalized boundary parameters, respectively.

To cover parabolic and elliptic monodromies, one has to
investigate analytical properties (in the variables $p$, $l_b$,
$r_b$) of the vertex operator $V$. Work in this direction is in
progress.

\section{Conclusions}

For Minkowski space Liouville theory on the strip we have
performed a complete analysis of classical solutions regular in
the bulk of the strip. These solutions, falling into conformal
co\-adjoint orbits of the energy-momentum tensor \cite{Balog}, can
be parameterized by the constant energy density $T_0$ of the
lowest energy solution in the orbit and an element $\xi (x)$ of
the conformal group of the strip.

Depending on the parameters $l$ and $r$, describing the
conformally invariant generalized Neumann boundary conditions
(FZZT branes) on the left and right boundary of the strip, the
solutions have elliptic, parabolic or hyperbolic monodromies.
Avoiding singularities in the bulk requires $l,r\geq -1$.
Solutions with elliptic monodromy correspond to bound states,
those with hyperbolic monodromy to scattering states. For $l+r>0$
all positive values of $T_0$ are allowed, the monodromy is then
always hyperbolic. For $l+r<0$ negative energies  above a
threshold depending on $l,r$ and elliptic monodromy are allowed as
well as all positive energies and hyperbolic monodromy. The
peculiarities of zero energy and parabolic monodromy have been
touched, too.

For $l$ or $r=-1$ and certain related $T_0$ the Liouville f\/ield
develops a controlled singularity on the boundaries, just
realizing a Dirichlet condition (ZZ brane).

For the Hamilton description of the system the Poisson brackets
and the canonical two form has been expressed in terms of the
variables $T_0$ and $\xi(x)$. To prepare the system for
quantization an alternative description in terms of a free f\/ield
has been given, similar to the corresponding construction for the
Liouville f\/ield theory on a cylinder \cite{BCT,BCT+, OW, T1,
JW}.

We could get a f\/irst estimate of quantum ef\/fects by discussing
semi-classical Bohr--Sommerfeld quantization. The bound state
energy levels become quantized and the spectrum agrees with the
corresponding quasiclassical limit of the spectrum gained
in~\cite{T} by conformal bootstrap techniques in Euclidean space.

Finally we have constructed the quantum version of the degenerated
exponential of the Liou\-ville f\/ield $ e^{-\varphi} $.  The
quantum deformation of the weights in its representation in terms
of free f\/ield exponentials has been f\/ixed by requiring
locality and hermiticity.

There is an obvious schedule for further investigations. From the
free f\/ield representation of $e^{-\varphi} $ in the hyperbolic
sector one can read of\/f the ref\/lection amplitude. Its poles
should give information on the full quantum bound state spectrum.
With the quantum $e^{-\varphi} $ at hand one can construct generic
correlation functions following the technique used for the
periodic case \cite{JW1}. We also hope to fully explore the
limiting ZZ case within the canonical quantization.

\appendix

\section{Calculation of 2-forms}

\subsection{Dirichlet condition}

Equation (\ref{V-xi-ZZ}) leads to the following parametrization of
the canonical coordinates
\begin{gather*}
\varphi(\tau,\sigma)= \frac{1}{2}\log{\xi'(x)\xi'(\bar
x)}-\log\sin\frac{1}{2}[\xi(x) -\xi(\bar x)]-\log 2m,
\\
\pi(\tau,\sigma) =\frac{\xi''(x)}{2\xi'(x)} +\frac{\xi''(\bar
x)}{2\xi'(\bar x)} -\frac{\xi'(x)-\xi'(\bar x)}{2}
\cot\frac{1}{2}[\xi(x)-\xi(\bar x)].
\end{gather*}
The canonical form (\ref{omega-can}) can be represented in the
form $\omega=\omega_0+\bar\omega_0+\omega_1$, where
\begin{gather*}
\omega_0=\frac{1}{4}\int_0^{\pi}d\sigma \left[
\frac{\delta\xi''(x)\wedge\delta\xi'(x)}{\xi'^2(x)}
-\delta\xi'(x)\wedge\delta\xi(x)\right],
\\
\bar\omega_0=\frac{1}{4}\int_0^{\pi}d\sigma\,\left[
\frac{\delta\xi''(\bar x)\wedge\delta\xi'(\bar x)}{\xi'\,^2(\bar
x)} -\delta\xi'(\bar x)\wedge\delta\xi(\bar x)\right],
\end{gather*}
while $\omega_1$ turns to a boundary term, since it is represented
as an integral from a derivative by~$\sigma$. The 2-form
$\omega_1$ vanishes due to the monodromy properties of $\xi$.
Using the doubling trick as in~(\ref{E}), we rewrite the sum
$\omega_0+\bar\omega_0$ into (\ref{omega-ZZ}).

\subsection{Neumann conditions}

The general solution (\ref{V-xi}) can be written in
 the standard Liouville form
\begin{equation*}
V=\frac{1+m^2 F(x)\bar F(\bar x)}{\sqrt{F'(x)\bar{F}'(\bar x)}},
\end{equation*}
with
\begin{equation}\label{A-xi,p}
F(x)=\frac{u e^{p\xi(x)}+le^{\pi p}+r}{2\sinh\pi p},\qquad \bar
F(\bar x)=\frac{u e^{p\xi(x)}+le^{\pi p}+r}{2\sinh\pi p}.
\end{equation}
Applying the same technique as before, we express the  canonical
form (\ref{omega-can}) in terms of parame\-terizing $F$ and $\bar F$
f\/ields
\begin{equation*}
\omega=\frac{1}{4}\int_0^{\pi}d\sigma \left[ \frac{\delta
F''(x)\wedge\delta F'(x)}{F'^2(x)} +\frac{\delta \bar F''(\bar
x)\wedge\delta \bar F'(\bar x)} {\bar F'^2(\bar
x)}\right]+\mbox{B.T}.
\end{equation*}
with the boundary term
\begin{gather*}
\mbox{B.T.}=\frac{\delta F'(x)\wedge\delta \bar F'(\bar
x)}{4F'(x)\bar F'(\bar x)}
-\frac{1}{2}\delta\log\!\left(\frac{F'(x)}{\bar F'(\bar
x)}\right)\!\wedge \delta\log(1+m^2 F(x)\bar F(\bar
x))+\frac{\delta F(x)\wedge\delta\bar F(\bar x)}{1+m^2F(x)\bar
F(\bar x)}
\Bigg|_0^\pi\!.
\end{gather*}
Then, using (\ref{A-xi,p}) and the monodromy properties of
$\xi$-f\/ield we get (\ref{omega-FZZT}).

\section{Comparison of quasiclassical quantization\\ with the
corresponding limit of the conformal\\ bootstrap spectrum}

First we write our formula (\ref{theta-0}) for $\theta _*(l,r)$ in
a form more suitable for the comparison with~\cite{T}. Denoting
$l=l_1$, $r=l_2$ and def\/ining $\vartheta _j$ in $(0,\pi )$ for
$\vert l_j\vert <1$ by
\begin{equation*}
l_j=\cos \vartheta _j
\end{equation*}
we get
\begin{equation*}
\theta _*(l,r)=\frac{\vartheta _1+\vartheta _2}{\pi}-1.
\end{equation*}
This brings (\ref{semi}) in the form
\begin{equation}\label{T-var}
(T_0)_n= -\frac{\hbar ^2}{4\pi^2}
 n^2+\frac{\hbar}{2\pi}\left (\frac{\vartheta_1+\vartheta_2}{\pi}-
1\right ) n- \frac{(\vartheta_1+\vartheta_2)^2 }{4\pi ^2} +
\frac{\vartheta _1+\vartheta
  _2}{2\pi} - \frac{1}{4}.
\end{equation}
The dictionary to compare our normalizations of the Liouville
f\/ield, the mass and boundary parameters $\varphi$, $m$, $l_j$
with that of \cite{T} ($\phi_T,\mu ,\rho _j$) is
\begin{equation*}
\varphi =b\phi _T,\qquad m^2=\mu\pi b^2,\qquad
l_j=\sqrt{\frac{\pi}{\mu}} b\rho _j.
\end{equation*}

According to  \cite{T}, the state space of Liouville theory on the
strip is the direct sum of highest weight ($\Delta _{\beta}=\beta
(Q-\beta ), Q=1/b+b$) representations of the Virasoro algebra.
There is a~continuum contribution $\beta\in Q/2+i\mathbb{R} ^+$,
and depending on the boundary parameters a discrete contribution
\cite{T} characterized by
\begin{equation}\label{T-spec}
\beta = Q-\vert \sigma _{\pm}\vert +nb+\hat
n\frac{1}{b}<\frac{Q}{2},
\end{equation}
where $n$, $\hat n$ are non-negative integers and $\sigma
_{\pm}=i(s_2\pm s_1)$ with
\begin{equation}\label{T-b}
\cosh (2\pi bs_j)=\frac{\rho _j}{\sqrt \mu}\sqrt {\sin (\pi b^2)}.
\end{equation}
The evaluation of (\ref{T-b}) in the quasiclassical limit
$b\rightarrow 0$ expressed in our boundary parameters~$l_j$ (for
$\vert l_j\vert <1$) gives
\begin{equation}\label{sigma-l}
s_j=i\frac{\arccos l_j}{2\pi b}+{\cal O}(b^2).
\end{equation}
Inserting this into (\ref{T-spec}) one f\/irst notices that for
small enough $b$ the option $\hat n \neq 0$ is switched of\/f. On
top of this, in this limit only the choices $\sigma _+$ and
$\arccos l_j\in (0,\pi)$ obey the inequality in~(\ref{T-spec}).
Altogether this leads to
\begin{equation*}
b^2\Delta _n=- b^4n(n+1)+b^2 \left
  (\frac{\vartheta_1+\vartheta_2}{\pi}-1\right
)n-\frac{(\vartheta_1+\vartheta_2)^2 }{4\pi ^2}+ \frac{\vartheta
_1+\vartheta
  _2}{2\pi}.
\end{equation*}
After the identif\/ications $\hbar =2\pi b^2$  and
$(T_0)_n=b^2\Delta _n$ this agrees with (\ref{T-var}) up to the
trivial shift by $-1/4$ and the replacement $n(n+1)\mapsto n^2$,
valid for large $n$ and common for the quasiclassical
approximation.  The continuous spectrum in \cite{T} corresponds to
the our solutions with hyperbolic monodromy.

Although not touching the issue of quantization for the Dirichlet
case in this paper, we nevertheless can add already one
interesting observation concerning the spectrum of $T_0$. From
Subsection~2.1 we know that classically there is only one value
for $T_0$ allowed. It is $T_0=-1/4$, if on both sides of the strip
Dirichlet conditions are imposed, and  $T_0=-(\arccos r)^2/(4\pi
^2)$ for Dirichlet on the left and generalized Neumann with
parameter $r$ on the right.  With the just derived translation
rule $T_0=b^2\Delta -1/4$ this corresponds to the conformal
dimensions of highest weight states of the contributing Verma
modules $\Delta = 0$ and $\Delta = s^2+1/(4b^2)$, respectively.
This agrees in leading order of $b$ with the full quantum result
via conformal bootstrap reported in~\cite{ZZ,Nak} for the $(1,1)$
ZZ brane. Note that $s_j$ def\/ined according to~\cite{T} in our
equation (\ref{sigma-l}) dif\/fers by a factor $1/2$ from $s$
in~\cite{ZZ,Nak}.

\section{Exchange relations}

\subsection[Poisson brackets algebra of chiral fields]{Poisson brackets algebra of chiral f\/ields}

In this appendix we use the method applied in \cite{FJ}. The
chiral f\/ield $\psi(x)=e^{-\phi(x)}$ is the classical analog of
the operator (\ref{psi^}) and the canonical Poisson brackets
(\ref{PB,a_n}) are equivalent to
\begin{eqnarray}\label{psi-psi}
\{\psi(x), \psi(y)\}=  \frac{1}{4}\epsilon(x-y)\psi(x)\psi(y),
\end{eqnarray}
which quantum mechanically becomes (\ref{psi-psi^}).

The operator (\ref{A^}) corresponds to the f\/ield
$A_p(x)=\int_0^{2\pi}dz e^{2\phi(y+z)-\pi p}$  and its Poisson
bracket with the $\psi$-f\/ield reads
\begin{eqnarray}\label{psi,A}
\{\psi(x), A_p(y)\}= - \frac{1}{2}\psi(x) \int_0^{2\pi}dz
e^{2\phi(y+z)-\pi p}(\epsilon(x-y-z)+1).
\end{eqnarray}
Due to the stair-step character of the $\epsilon$-function the
following identity holds
\begin{equation}\label{epsilon3}
\epsilon(a+b)-\epsilon (a)-\epsilon(b)=\pm 1,
\end{equation}
and since $\cosh \pi p\pm\sinh\pi p=e^{\pm\pi p}$, we f\/ind
\begin{equation}\label{eps-sh}
\epsilon(x-y-z)=\epsilon(x-y)-\epsilon(z)- \frac{\cosh\pi
p}{\sinh\pi p}+\frac{e^{\pi p[\epsilon(x-y-z)-
\epsilon(x-y)+\epsilon(z)]}}{\sinh\pi p}.
\end{equation}
Inserting (\ref{eps-sh}) into (\ref{psi,A}) and using that the
function $2\phi(y+z)+\pi p\epsilon(x-y-z)$ is periodic in~$z$, we
can shift the integration domain in the last term and obtain
\begin{gather}
\{\psi(x), A_p(y)\}= \frac{1}{2}\left( \frac{\cosh\pi p}{\sinh\pi
p}-\epsilon(x-y)\right)\psi(x) A_p(y) -\frac{e^{-\pi
p\epsilon(x-y)}}{2\sinh\pi p} \psi(x)A_p(x).\label{psi,A1}
\end{gather}

By (\ref{psi-psi}) and (\ref{psi,A1}) the f\/ield
$\chi(x)=\psi(x)A_p(x)$ satisf\/ies the relation
\begin{gather*}
\{\psi(x), \chi(y)\}= \frac{1}{2}\left( \frac{\cosh\pi p}{\sinh\pi
p}- \frac{1}{2} \epsilon(x-y)\right) \psi(x)\,\chi(y)
-\frac{e^{-\pi p\epsilon(x-y)}}{2\sinh\pi p}
\chi(x)\psi(y).
\end{gather*}
To f\/ind a closed form of the Poisson brackets
\begin{gather}\label{A,A}
\{A_p(x), A_p(y)\}= \int_0^{2\pi}dz e^{2\phi(x+z)-\pi p}
e^{2\phi(y+z')-\pi p} \epsilon(x-y+z-z')
\end{gather}
in terms of the $A_p$-f\/ield, we use the identity
\begin{gather}
\epsilon(x-y+z-z')=\epsilon(x-y)+\epsilon(z)-\epsilon(z')\nonumber\\
\qquad{}+ \frac{e^{-\pi p \epsilon(x-y)}}{\sinh\pi p} e^{\pi p
[\epsilon(x-y-z')+\epsilon(z')]}-\frac{e^{\pi
p\,\epsilon(x-y)}}{2\sinh\pi p}
e^{-\pi p\,[\epsilon(x-y+z)-\epsilon(z)]}\nonumber\\
\qquad{}+\frac{e^{\pi p \epsilon(z')}}{2\sinh\pi p} e^{\pi p
[\epsilon(x-y+z-z')-\epsilon(x-y+z)]}-\frac{e^{\pi
p\,\epsilon(z)}}{2\sinh\pi p} e^{-\pi p
[\epsilon(x-y+z-z')-\epsilon(x-y-z')]},\label{eps-sh1}
\end{gather}
which follows from (\ref{eps-sh}). The contributions of the last
two terms of (\ref{eps-sh1}) in the integral (\ref{A,A}) cancel
each other and provide the result
\begin{gather*}
\{A_p(x), A_p(y)\}= \epsilon(x-y) A_p(x)A_p(y)+\frac{e^{-\pi p
\epsilon(x-y)}}{2\sinh\pi p} A_p^2(x)-\frac{e^{\pi p
\epsilon(x-y)}}{2\sinh\pi p} A_p^2(y).
\end{gather*}
The calculation of the Poisson brackets between $\chi$-f\/ields is
now straightforward and we end up with
\begin{gather*}
\{\chi(x), \chi(y)\}=  \frac{1}{4} \epsilon(x-y) \chi(x) \chi(y),
\end{gather*}
which indicates that the f\/ields $\psi$ and $\chi$ are related
canonically.

\subsection{Operator algebra}

First note that the exchange  relations of $q$-exponentials and
$p$-dependent functions  is
\begin{gather}\label{psi-p^}
e^{aq} f(p)= f(p+ia\hbar)e^{aq}.
\end{gather}
An intermediate step towards the exchange relations between the
$\psi$ and $\chi$ operators is a calculation of the quantum analog
of (\ref{psi,A1}).
 Due to (\ref{phi-phi^}) and (\ref{psi-p^}),
from (\ref{psi^})--(\ref{A^}) we have
\begin{gather}\label{psi-A^}
\psi(x)A_p(y)= \int_0^{2\pi}dz e^{2\phi(y+z)}\, e^{-(\pi
p-i\hbar)} \psi(x)  e^{i(\hbar/2)\epsilon(x-y-z)}.
\end{gather}
To rewrite this equation as an exchange relation, we use the
identity
\begin{gather}
\sinh\pi p e^{i(\hbar/2)[\epsilon(x-y-z)
-\epsilon(x-y)+\epsilon(z)]}\nonumber\\
\qquad =\sinh(\pi p-i\hbar/2)+ i\sin(\hbar/2) e^{\pi
p[\epsilon(x-y-z)-\epsilon(x-y)+\epsilon(z)]},\label{eps-sh2}
\end{gather}
based on (\ref{epsilon3}). Inserting
$e^{i(\hbar/2)\epsilon(x-y-z)}$ from (\ref{eps-sh2}) into
(\ref{psi-A^}) we obtain
\begin{gather*}
\psi(x) A_p(y)=e^{{i(\hbar}/{2}) \epsilon(x-y)} \frac{\sinh\pi
p}{\sinh\left(\pi p+{i\hbar}/{2}\right)}
A_p(y) \psi(x)\nonumber\\
\phantom{\psi(x) A_p(y)=} {}+i\sin({\hbar}/{2})\frac{e^{-\pi
p\epsilon(x-y)}}{\sinh\left(\pi p+{i\hbar}/{2}\right)}
A_p(x)\psi(x),
\end{gather*}
which for $x=y$  yields
\begin{equation*}
\psi(x)A_p(x)=A_p(x) \psi(x).
\end{equation*}
The derivation of the exchange relation between the $\psi$ and
$\chi$ operators (see (\ref{chi^})) is now straightforward and we
obtain
\begin{gather}
\psi(x) \chi(y)= e^{{i(\hbar}/{4}) \epsilon(x-y)} \!\!
 \left[ \frac{\sinh\left(\pi
p-{i\hbar}/{2}\right)} {\sinh\pi p} \chi(y) \psi(x)
\! +\!i\sin({\hbar}/{2}) 
\frac{e^{-\pi p\epsilon(x-y)}} {\sinh \pi p}
 \psi(y) \chi(x)\right]\!\!.\label{psi-chi^}\!\!
\end{gather}
The exchange relation (\ref{chi-psi^}) is derived in a similar way
and (\ref{chi-psi^1}) follows from (\ref{psi-chi^})
and~(\ref{chi-psi^}) by simple algebraic manipulations.

The next step is the exchange relation between the
$A_p$-operators, which is obtained in the same manner and in a
symmetrized form it reads
\begin{gather*}
A_p(x) A_p(y) e^{{-i(\hbar}/{2}) \epsilon(x-y)}-A_p(y) A_p(x)
e^{{i(\hbar}/{2}) \epsilon(x-y)}\nonumber\\
\qquad{}=i\sin({\hbar}/{2})\left( \frac{e^{(\pi
p+i\hbar)\epsilon(x-y)}} {\sinh(\pi p+i\hbar)} A_p^2(y)-
\frac{e^{-(\pi p+i\hbar)\epsilon(x-y)}} {\sinh(\pi p+i\hbar)}
 A_p^2(x)\right).
\end{gather*}
This f\/inally provides (\ref{chi-chi^}).

\section[Locality and Hermiticity of $V$-operator]{Locality and Hermiticity of $\boldsymbol{V}$-operator}

The locality condition (\ref{local-herm}) is equivalent to the
symmetry of the product $V(\sigma, -\sigma)\,V(\sigma', -\sigma')$
under  $\sigma \leftrightarrow \sigma'$. Let us collect the terms
with a given power $N$ of the $\chi$-f\/ield. The number $N$
changes from 0 to 4. There is only one term with $N=0$
\begin{equation*}
C_{\sigma,\sigma'}=e^{-i\hbar/4}
\psi(-\sigma)\psi(\sigma)\psi(-\sigma')\psi(\sigma'),
\end{equation*}
which is symmetric due to (\ref{psi-psi^}). The case $N=4$ is
similar because of (\ref{chi-chi^}).

For the terms with $N=1$ we use the exchange relation
(\ref{chi-psi^1}), moving the $\chi$-f\/ield in each term to the
right hand side. Replacing then $\chi$ by $\psi A_p$, we f\/ind
the following structure
\begin{gather}\label{N=1}
\Lambda^{(1)}_pC_{\sigma,\sigma'} A_p(\sigma)+
\Lambda^{(2)}_pC_{\sigma,\sigma'} A_p(-\sigma)+
\Lambda^{(3)}_pC_{\sigma,\sigma'} A_p(\sigma')+
\Lambda^{(4)}_pC_{\sigma,\sigma'} A_p(-\sigma').
\end{gather}
with $p$ dependent coef\/f\/icients
$\Lambda_p^{(1)},\dots,\Lambda_p^{(4)}$. The symmetry of
(\ref{N=1}) requires $\Lambda^{(1)}_p =\Lambda^{(3)}_p$ and
$\Lambda^{(2)}_p=\Lambda^{(4)}_p$. These conditions lead to the
equations
\begin{gather*}
B_{p-i\hbar/\pi}= \frac{e^{i(\hbar/2)}}{\sinh(\pi p-i\hbar)}\left[
\frac{\sinh\left(\pi p-{i\hbar}/{2}\right)} {\sinh(\pi
p-3i\hbar/2)}B_p+i \frac{e^{-\pi p }\sin({\hbar}/{2})} {\sinh(\pi
p-3i\hbar/2)} C_p\right],
\\
C_{p-i\hbar/\pi}= i \frac{\sin({\hbar}/{2})e^{\pi p-i\hbar/2 }}
{\sinh(\pi p-i\hbar)} B_p+ \frac{e^{{-i(\hbar}/{2})}\sinh\left(\pi
p+{i\hbar}/{2}\right)} {\sinh(\pi p-i\hbar)}C_p,
\end{gather*}
which are simplif\/ied for the linear combinations
\begin{gather*}
X_p=\sinh\left(\pi p+{i\hbar}/{2}\right)C_p-
\sinh\left(\pi p-{i\hbar}/{2}\right)B_p,\\
\nonumber Y_p=-e^{-\pi p}\sinh\left(\pi p+{i\hbar}/{2}\right)C_p+
e^{\pi p}\sinh\left(\pi p-{i\hbar}/{2}\right)B_p,
\end{gather*}
in the form
\begin{gather*}
X_{p-i\hbar/\pi}=X_p,\qquad Y_{p-i\hbar/\pi}=Y_p.
\end{gather*}
Thus, with $X_p=2L$ and $Y_p=2R$, where $L$ and $R$ are arbitrary
complex numbers, we f\/ind
\begin{gather*}
B_p=\frac{L e^{-\pi p}+R}{\sinh\pi p \sinh(\pi p-i\hbar/2)},\qquad
C_p=\frac{L e^{\pi p}+R}{\sinh\pi p \sinh(\pi p+i\hbar/2)}.
\end{gather*}
The hermiticity condition (\ref{local-herm}) puts restrictions on
the parameters $L$ and $R$. Making use of the exchange relations
(\ref{psi-chi^}) and (\ref{chi-psi^}), one f\/inds a relation
between $B_p$ and $C_p$ and their complex conjugates, reducing the
freedom of two complex parameters to two real ones. With an
additional free real parameter from $D_p$ we f\/inally obtain with
real $l_b$, $r_b$ and $m_b$ (\ref{b_p}) and~(\ref{c_p}).

Due to the symmetry between the $\psi$ and $\chi$ f\/ields, the
case $N=3$ gives the same result as $N=1$.

The analysis of the case $N=2$ can be done similarly, but now with
the known $B_p$, $C_p$ and for $D_p$ we end up with (\ref{d_p}).

\subsection*{Acknowledgements}

We thank Cosmas Zachos for helpful discussions. G.J.\ is grateful to
the organizers of ``The O'Raifeartaigh Symposium'' for the
invitation. He thanks Humboldt University, AEI Golm, ICTP Trieste
and ANL Argonne for hospitality, where a main part of his work was
done. His research was supported by grants from the DFG (436 GEO
17/3/06) and GRDF (GEP1-3327-TB-03). H.D. was supported in part by
DFG with the grant DO 447-3/3.

\pdfbookmark[1]{References}{ref}
\LastPageEnding

\end{document}